\documentclass[11pt, oneside]{article}   	
\usepackage{geometry}                		
\usepackage{graphicx}				
\usepackage{amssymb}
\usepackage{amsmath}
\usepackage{subfig}
\usepackage{color}
\usepackage{graphicx}

\usepackage{multirow}
\usepackage{adjustbox}
\usepackage{setspace}
\usepackage{natbib}
\setcitestyle{authoryear,open={(},close={)}}

\newtheorem{theorem}{Result}

\makeatletter
\newcommand*{\indep}{%
	\mathbin{%
		\mathpalette{\@indep}{}%
	}%
}
\newcommand*{\nindep}{%
	\mathbin{
		\mathpalette{\@indep}{\not}
	}%
}
\newcommand*{\@indep}[2]{%
	\sbox0{$#1\perp\m@th$}
	\sbox2{$#1=$}
	\sbox4{$#1\vcenter{}$}
	\rlap{\copy0}
	\dimen@=\dimexpr\ht2-\ht4-.2pt\relax
	\kern\dimen@
	{#2}%
	\kern\dimen@
	\copy0 
} 
\makeatother

\usepackage{tikz}
\usetikzlibrary{arrows,shapes.arrows,shapes.geometric,backgrounds,decorations.pathmorphing,positioning,fit,automata}

\tikzset{
	>=stealth',
	true/.style={
		rectangle,
		draw=black, very thick,
		text width=6.5em,
		minimum height=2em,
		text centered,
		fill=gray, opacity = 0.5},
	punkt/.style={
		rectangle,
		rounded corners,
		draw=black, very thick,
		text width=6.5em,
		minimum height=2em,
		text centered},
	est/.style={
		circle,
		draw=black, very thick,
		text centered},
	shade/.style={
		circle,
		draw=black, very thick, fill=gray!50,
		text centered},
	weight/.style={
		circle,
		draw=black, very thick,
		text width=6.5em,
		minimum height=2em,
		text centered},
	pil/.style={
		->,
		thick,
		shorten <=2pt,
		shorten >=2pt,},
	double/.style={
		<->,
		thick,
		shorten <=2pt,
		shorten >=2pt,},
	dash/.style={
		dashed,
		thick,
		shorten <=2pt,
		shorten >=2pt,},
	dashdouble/.style={
		<->,
		dashed,
		thick,
		shorten <=2pt,
		shorten >=2pt}}

\title{Estimation of natural indirect effects robust to unmeasured confounding and mediator measurement error}
\date{}							

\author{Isabel R. Fulcher$^{1}$, Xu Shi$^{1}$, and Eric J. Tchetgen Tchetgen$^{2}$ \\ $^{1}$Department of Biostatistics, Harvard T.H. Chan School of Public Health \\ $^{2}$Wharton Statistics Department, University of Pennsylvania}

\begin{document}
	\maketitle
	\doublespacing

\begin{abstract}
	The use of causal mediation analysis to evaluate the pathways by which an exposure affects an outcome is widespread in the social and biomedical sciences. Recent advances in this area have established formal conditions for identification and estimation of natural direct and indirect effects. However, these conditions typically involve stringent no unmeasured confounding assumptions and that the mediator has been measured without error. These assumptions may fail to hold in practice where mediation methods are often applied. The goal of this paper is two-fold. First, we show that the natural indirect effect can in fact be identified in the presence of unmeasured exposure-outcome confounding provided there is no additive interaction between the mediator and unmeasured confounder(s). Second, we introduce a new estimator of the natural indirect effect that is robust to both classical measurement error of the mediator and unmeasured confounding of both exposure-outcome and mediator-outcome relations under certain no interaction assumptions. We provide formal proofs and a simulation study to demonstrate our results.  
\end{abstract}

\section{Introduction}  

Mediation analysis seeks to understand the underlying relationship between an exposure and outcome through an intermediate variable. That is, beyond evaluating the total effect of the exposure on outcome, one aims to evaluate the \textit{indirect} effect of the exposure on outcome through a given mediator and the \textit{direct} effect of the exposure on the outcome, not through the mediator. The seminal work of \cite{baron1986moderator} initially provided a regression framework for estimating these so-called indirect and direct effects. More recently, a formal counterfactual framework has provided a causal definition of these effects and corresponding conditions for identification and inference \citep{robins1992identifiability,pearl2001direct,imai2010general,vanderweele2015explanation,tchetgen2012semiparametric}. In this literature, the natural (pure) direct and indirect effects have emerged as the most common forms of mediation causal effects. Identification and estimation of these types of effects typically relies on the absence of unmeasured confounding and mediator measurement error. 

Much work in the causal inference literature has been devoted to addressing confounding bias when exposure total effect is the target parameter of primary interest. Concerns about unmeasured confounding can be considerably more severe in the mediation context as identification of natural direct and indirect effects have typically require no unmeasured confounding for the exposure-outcome, exposure-mediator, and mediator-outcome associations together with no exposure-induced confounding of the mediator, whether unmeasured or measured \citep{pearl2001direct, avin2005identifiability}. The assumption of no unmeasured confounding of the effects of exposure on the mediator and outcome in view is typically guaranteed by design in randomized experiments, whereby exposure assignment is under investigator control. However, bias due to unmeasured confounding of the mediator-outcome relation may be present even in randomized experiments as the mediator would typically not be under experimental control. To address this concern, sensitivity analysis can be used to produce a range of possible values for the effect estimate(s) as one varies assumptions about the degree of unmeasured confounding. However, these methods require additional assumptions about the nature of the relationship between the unmeasured confounders and exposure, mediator, or outcome variables and in the extreme give rise to bounds that may be too wide to be informative \citep{hafeman2011confounding,imai2010identification,vanderweele2014sensitivity,tchetgen2012semiparametric,tchetgen2014estimation}. Furthermore, in observational studies, all three sources of unmeasured confounding must be addressed, which makes such analyses both difficult to implement and interpret. 

The validity of estimators of both natural direct and indirect effects also relies on the mediating variable being measured without error. In the setting of a continuous outcome and mediator, it is possible to derive bias formulas for mediator subject to classical measurement error using standard measurement error theory \citep{carroll2006measurement}; see \cite{valeri2014mediation} for illustration. However, validation data is typically required to identify direct and indirect effects using these methods even under the fairly stringent parametric assumptions they impose on the observed data distribution. Unfortunately, validation data is rarely available in practice therefore making it impossible to identify the effects in view, and one is reduced to reporting sensitivity analyses often with no formal way of selecting meaningful values for sensitivity parameters.

The current paper addresses this gap in the literature by establishing conditions for point identification of natural indirect effects in the presence of exposure-outcome confounding, mediator-outcome confounding and measurement error of the mediator. We make two main contributions to the existing literature. First, we show that natural indirect effects can be identified under less stringent no confounding conditions than are typically assumed; that is, the no unmeasured confounding of the exposure-outcome relation can be replaced with the weaker assumption of no additive interaction between the mediator and unmeasured confounder(s) in the outcome model. As we show, this result in fact generalizes, and equally applies to continuous, binary, and censored failure time outcomes. The second contribution of the paper is to provide a new estimator for the natural indirect effect that is further robust to both unmeasured confounding of the mediator-outcome association and classical measurement error of the mediator under no additive interaction assumptions. An immediate implication of our results is that we show that no interaction assumptions made in prior sensitivity analyses by \cite{valeri2014mediation} can be strengthened by a slight additional assumption that variance of the mediator depends on exposure, in which case point identification is possible therefore obviating the need for sensitivity analysis. 

The remainder of this paper is organized as followed. In section 2, we present notation used throughout and identification conditions for natural direct and indirect effects used in current mediation literature focusing primarily on the natural indirect effect. In section 3, we describe a set of sufficient conditions for nonparametric identification of the natural indirect effect in the presence of exposure-outcome uncontrolled confounding. In section 4, we give identification conditions and propose estimators for the natural indirect effect in the presence of mediator-outcome unmeasured confounding (4.1), exposure-outcome unmeasured confounding (4.2), and classical measurement error of the mediator (4.3). Finally, in section 5, we provide a simulation study demonstrating our results.

\section{Notation} 

In the following, let $M(a)$ denote the counterfactual intermediate variable had the exposure taken value $a$ and $Y(a) = Y(a, M(a))$ denote the counterfactual outcome had exposure possibly contrary to the fact taken value $a$. We will also consider the counterfactual outcome $Y(a,M(a^*))$ had exposure taken its level $a$ and the intermediate variable taken the value it would have under $a^*$. Additionally, let $C$ be a set of observed pre-exposure covariates known to confound the $A$-$M$, $A$-$Y$ and $M$-$Y$ associations. Throughout $M$ can be vector valued. 

For exposure levels $a$ and $a^*$, the standard decomposition of the total causal effect into natural direct and indirect effects at the individual level is as follows,
$$ Y(a,M(a)) - Y(a^*,M(a^*)) =  \underbrace{Y(a,M(a)) - Y(a, M(a^*))}_\text{Natural Indirect Effect} + \underbrace{Y(a, M(a^*)) - Y(a^*,M(a^*))}_\text{Natural Direct Effect} $$
The natural indirect effect is the difference between the potential outcome under exposure value $a$ and the potential outcome if the exposure had taken value $a$ but the intermediate variable had taken the value it would have under $a^*$. At the population level, this is expressed upon taking expectation to obtain the average natural indirect effect,
\begin{align}
NIE = E[Y(a, M(a)) - Y(a,M(a^*))] \label{nie}
\end{align}
The natural direct effect is the difference between the potential outcome if the exposure had taken value $a$ but the intermediate variable had taken the value it would have under $a^*$ and the potential outcome under exposure value $a^*$. At the population level, this is expressed upon taking expectation to obtain the average natural indirect effect,
$$NDE = E[Y(a, M(a^*)) - Y(a^*,M(a^*))]$$ 
In the setting given in Figure 1, the NIE and NDE are typically identified under the following conditions: 
\begin{align*} 
\textrm{\textbf{M1.}} \ \ & \textrm{Consistency assumptions: } \textrm{(1) If $A=a$, then $M(a) =M$ w.p.1}, \\
& \hspace{4.8cm} \textrm{(2) If $A=a$, then $Y(a) =Y$ w.p.1}, \\
& \hspace{4.8cm} \textrm{(3) If $A=a$ and $M=m$, then $Y(a,m) =Y$ w.p.1} \\
\textrm{\textbf{M2.}}  \ \ & M(a^*) \perp A \mid C=c \ \ \ \forall \ a^*, c \\ 
\textrm{\textbf{M3.}} \ \ & Y(a,m) \perp  M(a^*) \mid A=a,C=c \ \ \ \forall \ m,a,a^*,c \\
\textrm{\textbf{M4.}} \ \ & Y(a,m) \perp  A \mid C=c \ \ \ \forall \ m,a, c 
\end{align*}
M1 is the consistency assumption, which states the observed variable is equal to the counterfactual variable corresponding to the observed intervention variable(s). M2 and M4 respectively state that there is no unmeasured confounding of the exposure and the intermediate variable and the exposure and the outcome. M3 encodes a no unmeasured confounding assumption of the mediator and the outcome, but is more stringent as it also implies independence between mediator and outcome potential values under conflicting treatment values. These conditions rule out the existence of unmeasured variables that confound the relationships between $A$-$M$, $A$-$Y$, and $M$-$Y$, and do not allow for exposure induced confounding of $M-Y$ association whether such variable is measured or not. 

Under assumptions M1-M4, 
\begin{align}
E[Y(a, M(a))] & = \int_c \int_m E[Y | M=m, A=a, C=c]\  f_M(m | a, c) \ f_C(c) \ dm \ dc \label{term1} \\
E[Y(a, M(a^*))] & = \int_c \int_m E[Y | M=m, A=a, C=c] \ f_M(m | a^*, c) \ f_C(c) \ dm \ dc  \label{term2}
\end{align}
such that 
\begin{align}
NIE(a,a^*)  & = \int_c \int_m E[Y | M=m, A=a, C=c]\  (f_M(m | a, c) - f_M(m | a^*, c)) \ f_C(c) \ dm \ dc \label{nie_identified1}
\end{align}

Expressions (\ref{term1})-(\ref{nie_identified1}) were first derived by \cite{pearl2012causal} under a nonparametric structural equation model with independent errors (NPSEM-IE) interpretation of the causal DAG in Figure 1. In the sections that follow, we show that assumptions M1-M4 can be modified to allow for some degree of uncontrolled confounding or mediator measurement error.

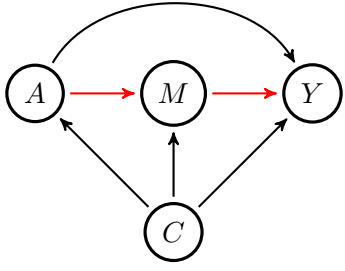
\begin{figure}
	\centering
		\begin{tikzpicture}[->,>=stealth',node distance=1cm, auto,]
		\node[est] (A) {$A$};
		\node[est, right = of A] (Z) {$M$};
		\node[est, right = of Z] (Y) {$Y$};
		\node[est, below = of Z] (C) {$C$};
		\path[pil,red] (A) edgenode {} (Z);
		\path[pil,red] (Z) edgenode {} (Y);
		\path[pil] (C) edgenode {} (A);
		\path[pil] (C) edgenode {} (Z);
		\path[pil] (C) edgenode {} (Y);
		\path[pil] (A) edge [bend left=60] node [left]  {} (Y);
		\end{tikzpicture} 
\caption{Directed Acyclic Graph (DAG) with indirect effect in red identified under a NPSEM-IE interpretation} \label{figure1}
\end{figure}

\begin{figure}
	\centering
	\subfloat[unmeasured confounding of $A$-$Y$ (violation of M4)]{
		\begin{tikzpicture}[->,>=stealth',node distance=1cm, auto,]
		\node[est] (A) {$A$};
		\node[est, right = of A] (Z) {$M$};
		\node[est, right = of Z] (Y) {$Y$};
		\node[shade, above = of Z] (W) {$W$};
		\path[pil,red] (A) edgenode {} (Z);
		\path[pil,red] (Z) edgenode {} (Y);
		\path[pil] (W) edgenode {} (A);
		\path[pil] (W) edgenode {} (Y);
		\path[pil] (A) edge [bend left=30] node [left]  {} (Y);
		\end{tikzpicture} } \hspace{1.5cm} 
	\subfloat[unmeasured confounding of $M$-$Y$ (violation of M3)]{
		\begin{tikzpicture}[->,>=stealth',node distance=1cm, auto,]
		\node[est] (A) {$A$};
		\node[est, right = of A] (Z) {$M$};
		\node[est, right = of Z] (Y) {$Y$};
		\node[shade, yshift=-1.8cm, xshift=2.8cm] (U) {$U$};
		\path[pil,red] (A) edgenode {} (Z);
		\path[pil,red] (Z) edgenode {} (Y);
		\path[pil] (U) edgenode {} (Z);
		\path[pil] (U) edgenode {} (Y);
		\path[pil] (A) edge [bend left=30] node [left]  {} (Y);
		\end{tikzpicture} } \\
	\subfloat[unmeasured confounding of $A$-$Y$ and $M$-$Y$ (violation of M3 and M4)]{
		\begin{tikzpicture}[->,>=stealth',node distance=1cm, auto,]
		\node[est] (A) {$A$};
		\node[est, right = of A] (Z) {$M$};
		\node[est, right = of Z] (Y) {$Y$};
		\node[shade, yshift=-1.8cm, xshift=2.8cm] (U) {$U$};
		\node[shade, above = of Z] (W) {$W$};
		\path[pil,red] (A) edgenode {} (Z);
		\path[pil,red] (Z) edgenode {} (Y);
		\path[pil] (U) edgenode {} (Z);
		\path[pil] (U) edgenode {} (Y);
		\path[pil] (W) edgenode {} (A);
		\path[pil] (W) edgenode {} (Y);
		\path[pil] (A) edge [bend left=30] node [left]  {} (Y);
		\end{tikzpicture} } \hspace{1.5cm} 
	\subfloat[measurement error of $M$ and unmeasured confounding of $A$-$Y$ and $M$-$Y$]{
		\begin{tikzpicture}[->,>=stealth',node distance=1cm, auto,]
		\node[est] (A) {$A$};
		\node[shade, right = of A] (Z) {$M$};
		\node[est, right = of Z] (Y) {$Y$};
		\node[shade, yshift=-1.8cm, xshift=2.8cm] (U) {$U$};
		\node[shade, above = of Z] (W) {$W$};
		\node[est, below = of Z] (Zs) {$M^*$};
		\path[pil,red] (A) edgenode {} (Z);
		\path[pil,red] (Z) edgenode {} (Y);
		\path[pil] (U) edgenode {} (Z);
		\path[pil] (Zs) edgenode {} (Z);
		\path[pil] (U) edgenode {} (Y);
		\path[pil] (W) edgenode {} (A);
		\path[pil] (W) edgenode {} (Y);
		\path[pil] (A) edge [bend left=30] node [left]  {} (Y);
		\end{tikzpicture} }
	\caption{Directed Acyclic Graphs (DAGs) with indirect effects in red. NIE under DAG (a), (b), (c), and (d) are discussed in Sections~\ref{2a}, \ref{2b}, \ref{2c}, and \ref{2d}, respectively. To ease exposition, the set of all covariates $C$ have been excluded from the graphs so that the graph can be viewed as conditional on $C$. See appendix for complete graphs. } \label{figure2}
\end{figure}
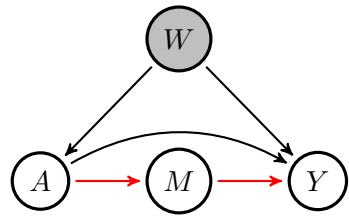
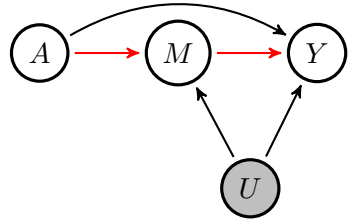
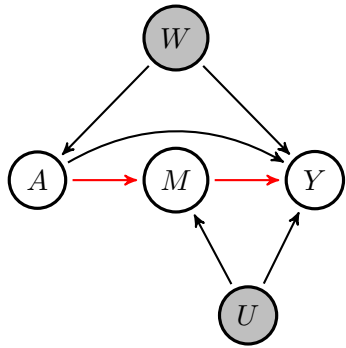
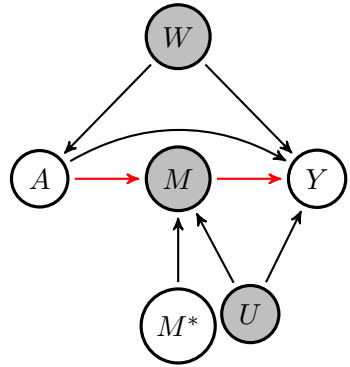

\section{Unmeasured confounding of the exposure-outcome relation} \label{2a}
Let $W$ denote a set of unobserved pre-exposure covariates that may confound the $A$-$Y$ association. We propose modifying M4 such that the NIE can be identified under DAG (a) in Figure 2. 

\textbf{M4$'$.}  There exists an unmeasured variable $W$ such that, 

\hspace{1cm}  (a) $ Y(a,m) \indep A \mid C=c, W=w \ \forall \ m,a,c,w $ 

\hspace{1cm}  (b) $M \indep W \mid C=c$ $\forall \ c$ 

\hspace{1cm}  (c) $ E[Y | M=m, A=a, W=w,C=c] - E[Y | M=0, A=a, W=w,C=c] = \gamma_1(a,m,c) $

\noindent The first two parts of this assumption state the existence of an unmeasured variable $W$ that confounds only the association between $A$ and $Y$ and has no direct effect on $M$. Part (c) of this assumption states that there is no additive interaction between the unmeasured confounder $W$ and the mediator $M$ in the outcome model. In other words, there is no dependence on $W$ of the causal effect of $M$ on $Y$ on the additive scale. Note that the assumption places no a priori restriction on the observed data distribution and is therefore not empirically testable without an additional assumption.

\begin{theorem}
	Under assumptions M1-M3 and M4$'$, the natural indirect effect is nonparametrically identified by equation (\ref{nie_identified1}).
\end{theorem}
Proofs of all results can be found in the appendix. Note that this result only pertains to the NIE contrast as the mean counterfactuals are themselves not nonparameterically identified under these conditions. Additionally, Result 1 extends to the conditional NIE given covariates. The result implies that unmeasured confounding of exposure-outcome relationship can largely be ignored when targeting NIE provided the unmeasured factor does not have a direct effect on $M$ and does not interact with $M$ (on additive scale). For inference, previously described parametric and semiparametric methods to target functional (\ref{nie_identified1}) can be used without modification \citep{baron1986moderator,pearl2001direct,vansteelandt2012natural,imai2010general,tchetgen2012semiparametric,valeri2013mediation,tchetgen2013inverse}.

\subsection{Binary outcomes} 

\noindent In the case of binary $Y$, one may instead wish to estimate the natural indirect effect on the multiplicative scale using the risk ratio, 
\begin{align}
RR^{NIE}(a,a^*|c) = Pr[Y(a, M(a))=1 | C=c] \big/ Pr[Y(a, M(a^*))=1 | C=c] \label{rr_nie}
\end{align}

The natural indirect effect on the risk ratio scale is identified under similar conditions to the NIE, namely, M4 can be replaced with the following assumption,
 
 \textbf{M4$''$.}  There exists an unmeasured variable $W$ such that, 
 
 \hspace{1cm}  (a) $ Y(a,m) \indep A \mid C=c, W=w \ \forall \ m,a,c,w $ 
 
 \hspace{1cm}  (b) $M \indep W \mid C=c$ $\forall \ c$ 
 
 \hspace{1cm}  (c) $ \frac{Pr[Y =1 | M=m, A=a, C=c, W=w]}{Pr[Y =1 | M=0, A=a, C=c, W=w]} = \gamma_2(a,m,c)$
 
\noindent The first two parts of assumption M4$''$ are identical to M4$'$, and part (c) now encodes no multiplicative interaction between the mediating variable and the unmeasured confounder. 

\begin{theorem}
	Under assumptions M1-M3 and M4$''$, the conditional natural indirect effect on the risk ratio scale is identified by 
	\begin{align}
	RR^{NIE}(a,a^*|c) & = \frac{\int_m Pr[Y=1 | M=m, A=a,C=c] f_M(m|a,c) dm }{\int_m Pr[Y=1 | M=m, A=a,C=c]  f_M(m|a^*,c) dm} \label{rr_nie_id}
	\end{align}
\end{theorem}

\noindent Similar to the result of Result 1, this result implies that unmeasured confounding of exposure-outcome relationship can largely be ignored when targeting the conditional NIE on the risk ratio scale provided the unmeasured factor does not have a direct on $M$ and does not interact with $M$ (on multiplicative scale). For estimation and inference, parametric and semiparametric methods to target functional (\ref{rr_nie_id}) have been developed by \cite{vanderweele2010bias} and \cite{tchetgen2013inverse}. 

\subsection{Other extensions and related literature} 
Further, mediation methods have been extended to include survival outcomes \citep{lange2011direct, tchetgen2011causal, vanderweele2011causal}. In the appendix, we show that the conditional natural indirect effect on the hazard difference scale remains identified under the analogous conditions as the natural indirect effect for the risk ratio scale. Additionally, we show that the conditional natural indirect effect on the hazard ratio scale is identified provided the outcome remains rare throughout follow-up. Therefore, existing statistical methods to estimate this quantity can directly be applied ignoring the advent of uncontrolled exposure-outcome confounding. 

There are two closely related strands of work in previous literature on mediation methods robust to exposure-outcome uncontrolled confounding. \cite{tchetgen2017evaluation} identified the NIE in the presence of unmeasured common cause of the exposure-outcome relation by replacing assumption M4 with an exclusion restriction, which states that the mediator potential outcome under no exposure is constant for all persons in the population. The exclusion restriction likely holds in the context of medication-mediated effects, where the exposure is typically disease status, mediator is medication taken for the disease, and outcome is an unintended effect of medication, so that healthy persons are usually excluded from receiving medication for the disease in question. In a separate strand of work,  \cite{fulcher2017generalized} propose a new form of indirect effect, the population intervention indirect effect (PIIE). This novel type of indirect effect captures the extent to which the effect of exposure is mediated by an intermediate variable under an intervention which fixes the component of exposure directly influencing the outcome at its observed value. More formally, 

$$ PIIE(a,a^*) = E[Y(A,Z(A))] - E[Y(Z(A,a^*))] = E[Y] - E[Y(Z(a^*))] $$

\noindent Unlike the NIE, the PIIE is nonparametrically identified under conditions M1-M3, therefore allowing for uncontrolled confounding of exposure-outcome relation without the need for an additional assumption such as no mediator-unmeasured confounding interaction. Additionally, the counterfactual mean $E[Y(Z(a^*))]$ is identified in the presence of exposure-outcome confounding, which is not the case for the counterfactual means in the NIE contrast. This result applies both conditional and unconditional on covariates.

\section{GENIUS for identification of the NIE} 

\subsection{Unmeasured confounding of the mediator-outcome relation}  \label{2b}

Let $U$ denote a set of unobserved pre-exposure covariates that may confound the $M$-$Y$ association.  We now consider the data generating mechanism represented by DAG (b) in Figure~\ref{figure2}. In this graph, the presence of the variable $U$ now subjects mediator-outcome to uncontrolled confounding; however, the exposure remains unconfounded (i.e. $W$ does not exist). Consistent with DAG (b), we consider sufficient conditions to identify the natural indirect effect,

\textbf{M3$'$.}  There exists an unmeasured variable $U$ such that, 

\hspace{1cm}  (a) $ Y(a,m) \indep M(a^*) \mid C=c, U=u \ \forall \ a,a^*,c,u $ 

\hspace{1cm}  (b) $A \indep U \mid C=c$ $\forall \ c$ 

\hspace{1cm}  (c) There is no additive $M$--($U,A,C$) interaction in the model for $E[Y | A, M, C, U]$ 

$$E[Y | A, M=m, C, U] - E[Y | A, M=0, C, U] = \theta_m m$$

\hspace{1.7cm} and no additive $A$--($U,C$) interaction in the model for $E[Y | A, M, C, U]$ 

$$E[Y | A=a, M, C, U] - E[Y | A=0, M, C, U] = \theta_a(a)$$

\hspace{1.7cm} for unknown function $\theta_a(.)$ that satisfies $\theta_a(0) =0$.

\hspace{1cm}  (d) There is no additive $A$--($U,C$) interaction in the model for $E[M | A, C, U]$ 

$$E[M | A=a, C, U] - E[Y | A=0, C, U] = \beta_a a $$


\hspace{1cm}  (e) $\text{var}(M|A=a,C=c)-\text{var}(M|A=a',C=c)\neq 0 \ \ \forall \ a, a', c$

The first two parts of this assumption state the existence of an unmeasured variable $U$ that confounds only the association between $M$ and $Y$ and has no direct effect on $A$.  Conditions (c)-(d) encode various no interaction assumptions in both the outcome and mediator models. Further, $C$ can in fact interact with $A$ and $M$ in the outcome model and $A$ in the mediator model; see appendix for more details as we have excluded here for ease of exposition. The last condition requires that $A$ must influence the variance of $M$ within each level of $C$. This is in fact a strengthening of the condition that the exposure is associated with the mediator after conditioning on $U$, which is expected to hold assuming the NIE is not null. This assumption is empirically testable for discrete $A$ and $C$ and will typically hold for binary $M$ except under certain exceptional laws \citep{tchetgen2017genius}. For continuous $M$, the result can be motivated by considering an underlying model for $M$ with latent heterogeneity in the effect of $A$ on $M$; see \cite{tchetgen2017genius} for further discussion. 

Under assumption M3$'$ parts (b) through (e), \cite{tchetgen2017genius} established that the average causal effect of $M$ on $Y$ is $\theta_m$ identified by, 
\begin{align}
\theta_m & = \frac{E\left[\{A-E[A|C]\}\{M-E[M|  A,C]\}Y\right]}{E\left[\{A-E[A|C]\}\{M-E[M|  A,C]\}M\right]}  \label{mrgenius1}
\end{align}

\cite{tchetgen2017genius} proposed equation (\ref{mrgenius1}) in the context of Mendellian Randomization (MR) estimation with an invalid instrumental variable (IV). In their setting, $A$ plays the role of an invalid IV as it violates the exclusion restriction of no direct effect on the outcome not through the endogenous variable $M$. They referred to this estimator as ``G-Estimation under No Interaction with Unmeasured Selection" (GENIUS) to reflect its close relationship to Robins G-estimation approach. Their result is related to the identification conditions given by \cite{lewbel2012using} under further linearity restrictions on models for $Y$ and $A$ although they establish generalization of GENIUS. We have the following result, 

\begin{theorem}
Under assumptions M1, M2, and M3$'$, the natural indirect effect is uniquely identified by 
\begin{align}
NIE(a,a^*) = \theta_m \beta_a(a - a^*) \label{nie_product}
\end{align}  
where $\theta_m$ is identified by equation (\ref{mrgenius1}) and $\beta_a$ is identified by standard regression of $M$ on $A$ and $C$.  
\end{theorem}

\subsection{Unmeasured confounding of exposure-outcome and mediator-outcome relations} \label{2c}
We now combine Result 1 and Result 3 to give identification conditions for the NIE that are robust to unmeasured confounding of both exposure-outcome and mediator-outcome associations. That is, both $U$ and $W$ are present. We propose modifying M3 and M4 such that the NIE can be identified under DAG (c) in Figure~\ref{figure2}, 

\textbf{M3$''$.}  There exists unmeasured variables $U$ and $W$ such that, 

\hspace{1cm}  (a) $ Y(a,m) \indep A \mid C=c, W=w \ \forall \ m,a,c,w $ 

\hspace{1cm}  (b) $A \indep U \mid C=c$ $\forall \ c$ 

\hspace{1cm}  (c) $ Y(a,m) \indep M(a^*) \mid C=c, U=u \ \forall \ a,a^*,c,u $ 

\hspace{1cm}  (d) $M \indep W \mid C=c$ $\forall \ c$ 

\hspace{1cm}  (e) $U \indep W \mid C=c$ $\forall \ c$ 

\hspace{1cm}  (f) There is no additive $M$--($U,W,A,C$) interaction in the model for $E[Y | A, M, C, U, W]$ 
$$E[Y | A, M=m, C, U, W] - E[Y | A, M=0, C, U, W] = \theta_m m$$
\hspace{1.7cm} and no additive $A$--($U,W,C$) interaction in the model for $E[Y | A, M, C, U, W]$ 
$$E[Y | A=a, M, C, U, W] - E[Y | A=0, M, C, U, W] = \theta_a(a)$$
\hspace{1.7cm} for unknown function $\theta_a(.)$ that satisfies $\theta_a(0) =0$.

\hspace{1cm}  (g) There is no additive $A$--($U,C$)interaction in the model for $E[M | A, U, C]$ 

$$E[M | A=a, C, U] - E[Y | A=0, C, U] = \beta_a a$$


\hspace{1cm}  (h) $\text{var}(M|A=a,C=c)-\text{var}(M|A=a',C=c)\neq 0 \ \ \forall \ (a,a')$ and all values of $c$

The first four parts of this assumption state the existence of unmeasured variables $U$ and $W$ that confound only the association $M-Y$ and $A-Y$, respectively. Additionally, part (d) encodes independence of the unmeasured variables given the observed set of covariates $C$. Similar to the assumptions given in M3$'$, parts (f) and (g) encode no interaction assumptions between certain variables in the model. 

\begin{theorem}
	Under assumptions M1, M2, and M3$''$, the natural indirect effect is uniquely identified by equation (\ref{nie_product}).
\end{theorem}

\subsection{Extension to allow for a mediator measurement error} \label{2d}
In this section, we extend the previous result and establish that the NIE remains uniquely identified under classical measurement error of the mediator, consistent with DAG (d) in Figure~\ref{figure2}.  Let $M^*$ denote a mismeasured version of the mediator $M$, under the classical measurement error model, 

\textbf{E1.} $M^* = M + \epsilon$ 

\textbf{E2.} $(A,Y,M,C,W,U)\perp \epsilon$

\textbf{E3.} $E[\epsilon]=0$

\noindent The above assumptions encode that the measurement error has mean zero and is independent of $A$, $Y$, $M$, $C$, $W$, and $U$. It is well known from measurement error literature, that under a linear model for the outcome where $M^*$ is used in place of $M$, the regression coefficient $\theta_m$ is not identified by standard least squares estimand and is biased towards the null \citep{carroll2006measurement}. In this case, the natural indirect effect is still given by (\ref{nie_product}). In addition to measurement error, we also allow for unmeasured confounding of $A$--$Y$ and $M$--$Y$ relations. We slightly modify M3$''$(h) by replacing $M$ with $M^*$,
 $$\text{var}(M^*|A=a,C=c)-\text{var}(M^*|A=a',C=c)\neq  0 \ \ \forall \ (a,a') \textrm{ and all values of } c$$
We refer to the updated set of conditions as M3$'''$ and obtain the result,

\begin{theorem}
	Under assumptions M1, M2, M3$'''$ and E1-E3, $\theta_m$ is identified by the following,  
	\begin{align*}
	\theta_m & = \frac{E\left[\{A-E[A|C]\}\{M^*-E[M^*|  A,C]\}Y\right]}{E\left[\{A-E[A|C]\}\{M^*-E[M^*|  A,C]\}M^*\right]}  
		\end{align*} 
\end{theorem}

\noindent Note that $\beta_a$ remains identified despite measurement error of the mediator. It follows that under the same set of assumptions, the NIE is uniquely identified by equation (\ref{nie_product}). 

\subsection{Estimation and software implementation of GENIUS} 
In the previous sections, we have provided identification conditions for the NIE in the presence of unmeasured confounding or measurement error of the mediator. Our identification formulae provide a straightforward approach to estimation by simply substituting unknown quantities with their empirical analogs. Specifically, the GENIUS estimator of $\theta_m$ can be consistently estimated by solving the following equation,
\begin{align}
0 & = n^{-1} \sum_{i=1}^n \bigg[ h(C_i) \big\{ A_i - \hat{E}[A | C_i] \big\} \big\{ M_i - \hat{E}[M | A_i, C_i] \big\} \big\{ Y_i - \hat{\theta}_mM_i \big\} \bigg] \label{mrgenius1_hat}
\end{align}

\noindent for user specified choice of $h$, where $\hat{E}[A | C]$ and $\hat{E}[M | A, C]$ are consistent estimators of $E[A | C]$ and $E[M | A, C]$ obtained by fitting generalized models. The \texttt{MR-GENIUS} R package provides an estimate for $\theta_m$ provided $C$ is empty \citep{mrgeniusgithub}. Note that heteroskedastic variance assumption is testable and empirical test is given by software.

The natural indirect effect estimator then follows from the product rule,
\begin{align}
\widehat{NIE}(a,a^*) = \hat{\theta}_m \hat{\beta}_a(a-a^*) \label{nie_genius}
\end{align}
\noindent where, as previously mentioned, the standard OLS estimator may be used for $\beta_a$ which can be implemented using any off-the-shelf statistics software package. 

For inference, the delta method or nonparametric bootstrap can be used to obtain variance estimates for the natural indirect effect. For the delta method, the form of the variance is given by,
$$ \widehat{Var}(\widehat{NIE}) = \hat{\theta}_m^2 s^2_{\hat{\theta}_m}+ \hat{\beta}_a^2 s^2_{\hat{\beta}_a}$$
where $s_{\hat{\theta}_m}$ is the standard error for $\hat{\theta}_m$ and can be extracted from the \texttt{MR-GENIUS} R package. And,  $s_{\hat{\beta}_a}$ is the standard error for $\hat{\beta}_a$ for which the robust sandwich variance should be used to account for heterogeneous variance \citep{white1980heteroskedasticity}. The nonparametric bootstrap may be preferable as the delta method may produce unstable estimates of the standard errors \citep{saunders2017classical, mackinnon1995simulation, mackinnon2004confidence, stone1990robustness}. 

\section{Simulation study}

\subsection{Simulation setup} 
We conduct extensive simulation studies under DAG (a)-(d). Our goal is to illustrate: 
\begin{enumerate}
	\item The estimator of the NIE given by the standard mediation formula is unbiased in the presence of unmeasured confounding of the exposure-outcome relationship under M1-M3 and M4$'$(Section 3, DAG (a))
	\item The proposed GENIUS estimator of the NIE is robust to unmeasured confounding of the mediator-outcome relationship, unmeasured confounding of the exposure-outcome relationship, and measurement error of the mediator and therefore delivers valid point and interval estimators (Section 4, DAG (b)-(d))
\end{enumerate}We simulate 2,000 data sets of sample size 1,000 under the following true data generating mechanism
\begin{itemize}
\item $W\sim \mathrm{Normal}(0,1)$ under DAG (a), (c), (d);
\item $U\sim \mathrm{Normal}(0,1)$ under DAG (b), (c), (d);
\item $A|W\sim \mathrm{Bernoulli}(p=\mathrm{expit}(w))$;
\item $M|A,U\sim \mathrm{Normal}(a+u, |0.5+0.5a|)$, while the observed mediator under DAG (d) is $M^*=M+\epsilon$ where $\epsilon\sim\mathrm{Normal}(0,1)$;
\item $Y|A,M,U,W\sim \mathrm{Normal}(a+m-u-w, 1)$.
\end{itemize}

Under these models, the natural indirect effect is equal to 1. We plan to use both naive ordinary least square (OLS) which fails to appropriately account for $U$, $W$, and measurement error, and the GENIUS estimator using the \texttt{MR-GENIUS} package. To provide a benchmark, we implemented the oracle estimator via OLS adjusting for $U$ and $W$ and error free mediator. 
Inference was performed using both the delta method and the nonparametric bootstrap. For the delta method, we used the Huber-White sandwich estimator for $\hat{\sigma}^2_{\beta_A}$ and $\hat{\sigma}^2_{\theta_m}$ was extracted from the \texttt{MR-GENIUS} package. For the bootstrap, we used 2,000 bootstrap samples with replacement.
Performance was evaluated in terms of Monte Carlo bias and variance of estimated NIE, the Monte Carlo mean of variance estimates, the proportion of bias, the mean squared error, and the coverage of the 95\% confidence interval using both the delta method and bootstrap variance estimates.

\subsection{Simulation results}
\begin{figure}[!htbp]
	\includegraphics[width=0.9\textwidth]{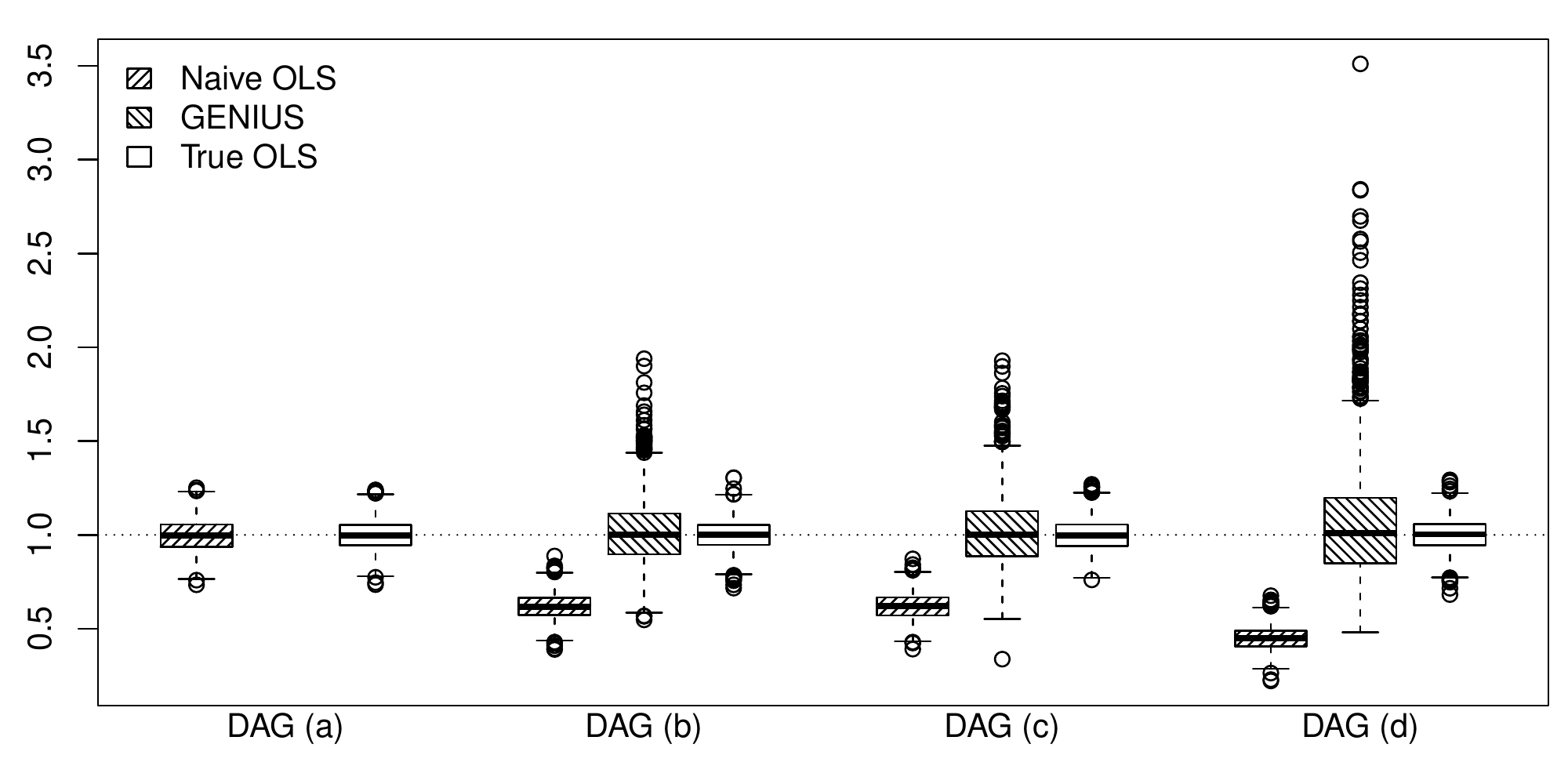}
	\caption{Boxplots of the estimates over 1000 samples using the naive OLS estimator, the GENIUS estimator, and the oracle OLS estimator under DAGs (a)-(d). True NIE is 1.}
	\label{simu_fig}
\end{figure}

Figure~\ref{simu_fig} and Table~\ref{simu_tbl} summarize simulation results for the naive OLS, the GENIUS, and the oracle estimators of the natural indirect effect. In presence of unmeasured confounding of the exposure-outcome relationship (DAG (a)), the naive OLS estimator has small bias and correct coverage and performs almost identically to the oracle estimator. For comparison, we also computed the GENIUS estimator, but if there is only unmeasured confounding of the exposure-outcome relation, its use is not recommended as it is less efficient. 

Under DAG (b) and (c), the GENIUS estimator has small bias and correct coverage, although it is slightly less efficient than the oracle estimator. When there also exists measurement error of the mediator (DAG (d)), the GENIUS approach is robust to mismeasured mediator with small proportion of bias. In contrast, the naive OLS estimator performs poorly in scenarios (b)-(d).  These findings confirmed that the proposed GENIUS estimator of the NIE is robust to unmeasured confounding of the mediator-outcome relationship, unmeasured confounding of the exposure-outcome relationship, and measurement error of the mediator with valid inference.

\begin{table}[!htp]
\caption{Operating characteristics of the naive OLS estimator, the GENIUS estimator, and the oracle estimator under directed acyclic graphs (DAGs) in Figure 2 (a)-(d). True NIE is 1.}\label{simu_tbl}
\begin{adjustbox}{center}
\begin{tabular}{clcccccc}
\hline 
\multirow{2}{*}{DAG} & \multirow{2}{*}{Method} & \multirow{2}{*}{Bias (Var)} & Proportion & Mean & Mean Var & 95\% CI & Bootstrap\tabularnewline
 &  &  & Bias (\%) & Squared error & Estimate & Coverage & Coverage\tabularnewline
\hline 
\multirow{3}{*}{(a)} & Naive OLS & -0.002 (0.007) & -0.2\% & 0.007 & 0.008 & 0.958 & 0.958\tabularnewline
 & GENIUS & -0.001 (0.014) & -0.1\% & 0.014 & 0.015 & 0.957 & 0.957\tabularnewline
 & Oracle & -0.002 (0.007) & -0.2\% & 0.007 & 0.007 & 0.959 & 0.958\tabularnewline
 &  &  &  &  &  &  & \tabularnewline
\multirow{3}{*}{(b)} & Naive OLS & -0.381 (0.005) & -38.1\% & 0.150 & 0.005 & 0.001 & 0.001\tabularnewline
 & GENIUS & 0.014 (0.030) & 1.4\% & 0.030 & 0.032 & 0.951 & 0.952\tabularnewline
 & Oracle & -0.000 (0.007) & -0.0\% & 0.007 & 0.007 & 0.952 & 0.953\tabularnewline
 &  &  &  &  &  &  & \tabularnewline
\multirow{3}{*}{(c)} & Naive OLS & -0.381 (0.005) & -38.1\% & 0.150 & 0.005 & 0.001 & 0.001\tabularnewline
 & GENIUS & 0.018 (0.036) & 1.8\% & 0.037 & 0.038 & 0.955 & 0.956\tabularnewline
 & Oracle & -0.000 (0.007) & -0.0\% & 0.007 & 0.007 & 0.948 & 0.947\tabularnewline
 &  &  &  &  &  &  & \tabularnewline
\multirow{3}{*}{(d)} & Naive OLS & -0.551 (0.004) & -55.1\% & 0.307 & 0.004 & 0.000 & 0.000\tabularnewline
 & GENIUS & 0.054 (0.091) & 5.4\% & 0.094 & 0.126 & 0.943 & 0.961\tabularnewline
 & Oracle & 0.001 (0.007) & 0.1\% & 0.007 & 0.007 & 0.946 & 0.943\tabularnewline
\hline 
\end{tabular}
\end{adjustbox}
\end{table}

\section{Discussion}

Unmeasured confounding and measurement error are two common sources of bias that can greatly impact estimation of natural indirect effects in mediation analysis. Existing methods to address these biases have primarily focused on conducting sensitivity analyses under various parametric assumptions about the mediator and outcome models. Further, there has been little focus on unmeasured confounding of the exposure-outcome relation, as it is often assumed the exposure is unconfounded. To this end, the first major contribution of our work shows that the no unmeasured confounding of the exposure-outcome relation for identification of the natural indirect effect can be replaced with the weaker assumption that there is no additive interaction between the mediator and unmeasured confounder(s) of the exposure-outcome relation. Conveniently, most of the estimators for the NIE available in standard software will be robust to unmeasured confounding of this form \citep{valeri2013mediation,valeri2015sas,tchetgen2013inverse,imai2010general,tchetgen2012semiparametric}. For investigators who do not wish to make the no interaction assumption, one may consider instead estimating a different form of indirect effect known as the population intervention indirect effect (PIIE) developed by \cite{fulcher2017generalized}. 

Second, our work also gives rise to a new estimator of the natural indirect effect that is robust to classical measurement error of the mediator and unmeasured confounding of both exposure-outcome and mediator-outcome relations. To the best of our knowledge, there is currently no estimator of the natural indirect effect that is robust to classical measurement error of the mediator without requiring additional information (i.e. validation data). Additionally, this is the first natural indirect effect estimator that is robust to unmeasured confounding of the mediator-outcome and exposure-outcome relation, which is especially a powerful result in randomized studies where this is the only potential source for confounding. Similar work includes \cite{vansteelandt2012natural}, which considered the indirect effect on the exposed in the presence of exposure-induced mediator outcome confounding, and \cite{zheng2015causal}, who propose an semiparametric estimation procedure for controlled effects that are robust to unmeasured confounding of the mediator-outcome relation. 

As noted throughout, our second result relies on parametric assumptions for both the outcome and mediator model. Although these parametric assumptions are common in the mediation literature, investigators should exercise caution when using these methods if the true data generating mechanism cannot reasonably follow from these assumptions. Standard procedures for assessing goodness of fit should be performed. When the assumptions made herein do not hold, one may still implement previously developed techniques for handling unmeasured confounding of the mediator-outcome \citep{le2016bias,vanderweele2010bias} and/or measurement error of the mediator \citep{le2012quantification,valeri2014mediation} for more general model specifications. 

\newpage 
	\singlespacing
\begin{center}
	\huge
	\textit{Supplementary Materials:} Estimation of natural indirect effects robust to unmeasured confounding and mediator measurement error
\end{center}

\section*{Alternate Figure 2}

\setcounter{figure}{3}
\begin{figure}[!htbp]
	\centering
	\subfloat[unmeasured confounding of $A$-$Y$ (violation of M4)]{
		\begin{tikzpicture}[->,>=stealth',node distance=1cm, auto,]
		\node[est] (A) {$A$};
		\node[est, right = of A] (Z) {$M$};
		\node[est, right = of Z] (Y) {$Y$};
		\node[shade, above = of Z] (W) {$W$};
		\node[est, below = of A] (C) {$C$};
		\path[pil,red] (A) edgenode {} (Z);
		\path[pil,red] (Z) edgenode {} (Y);
		\path[pil] (W) edgenode {} (A);
		\path[pil] (W) edgenode {} (Y);
		\path[pil] (A) edge [bend left=30] node [left]  {} (Y);
		\path[pil] (C) edge [bend left=70] node [left] {} (W);
		\path[pil] (C) edgenode {} (A);
		\path[pil] (C) edgenode {} (Z);
		\path[pil] (C) edgenode [left]  {} (Y);
		\end{tikzpicture} } \hspace{1.5cm} 
	\subfloat[unmeasured confounding of $M$-$Y$ (violation of M3)]{
		\begin{tikzpicture}[->,>=stealth',node distance=1cm, auto,]
		\node[est] (A) {$A$};
		\node[est, right = of A] (Z) {$M$};
		\node[est, right = of Z] (Y) {$Y$};
		\node[shade, yshift=-1.8cm, xshift=2.8cm] (U) {$U$};
		\node[est, below = of A] (C) {$C$};
		\path[pil,red] (A) edgenode {} (Z);
		\path[pil,red] (Z) edgenode {} (Y);
		\path[pil] (U) edgenode {} (Z);
		\path[pil] (U) edgenode {} (Y);
		\path[pil] (A) edge [bend left=30] node [left]  {} (Y);
		\path[pil] (C) edgenode {} (U);
		\path[pil] (C) edgenode {} (A);
		\path[pil] (C) edgenode {} (Z);
		\path[pil] (C) edgenode [left]  {} (Y);
		\end{tikzpicture} } \\
	\subfloat[unmeasured confounding of $A$-$Y$ and $M$-$Y$ (violation of M3 and M4)]{
		\begin{tikzpicture}[->,>=stealth',node distance=1cm, auto,]
		\node[est] (A) {$A$};
		\node[est, right = of A] (Z) {$M$};
		\node[est, right = of Z] (Y) {$Y$};
		\node[est, below = of A] (C) {$C$};
		\node[shade, yshift=-1.8cm, xshift=2.8cm] (U) {$U$};
		\node[shade, above = of Z] (W) {$W$};
		\path[pil,red] (A) edgenode {} (Z);
		\path[pil,red] (Z) edgenode {} (Y);
		\path[pil] (U) edgenode {} (Z);
		\path[pil] (U) edgenode {} (Y);
		\path[pil] (W) edgenode {} (A);
		\path[pil] (W) edgenode {} (Y);
		\path[pil] (A) edge [bend left=30] node [left]  {} (Y);
		\path[pil] (C) edgenode {} (U);
		\path[pil] (C) edge [bend left=70] node [left] {} (W);
		\path[pil] (C) edgenode {} (A);
		\path[pil] (C) edgenode {} (Z);
		\path[pil] (C) edgenode  {} (Y);
		\end{tikzpicture} } \hspace{1.5cm} 
	\subfloat[measurement error in $M$ and unmeasured confounding of $A$-$Y$ and $M$-$Y$]{
		\begin{tikzpicture}[->,>=stealth',node distance=1cm, auto,]
		\node[est] (A) {$A$};
		\node[shade, right = of A] (Z) {$M$};
		\node[est, right = of Z] (Y) {$Y$};
		\node[est, below = of A] (C) {$C$};
		\node[shade, yshift=-1.8cm, xshift=2.8cm] (U) {$U$};
		\node[shade, above = of Z] (W) {$W$};
		\node[est, below = of Z] (Zs) {$M^*$};
		\path[pil,red] (A) edgenode {} (Z);
		\path[pil,red] (Z) edgenode {} (Y);

		\path[pil] (U) edgenode {} (Z);
		\path[pil] (Zs) edgenode {} (Z);
		\path[pil] (U) edgenode {} (Y);
		\path[pil] (W) edgenode {} (A);
		\path[pil] (W) edgenode {} (Y);
		\path[pil] (A) edge [bend left=30] node [left]  {} (Y);
		\path[pil] (C) edge [bend left=-50] node [right] {} (U);
		\path[pil] (C) edge [bend left=70] node [left] {} (W);
		\path[pil] (C) edgenode {} (A);
		\path[pil] (C) edgenode {} (Z);
		\path[pil] (C) edgenode [left]  {} (Y);
		\end{tikzpicture} }
	\caption{Directed Acyclic Graphs (DAGs) with indirect effects in red. NIE under DAG (a), (b), (c), and (d) are discussed in Sections 3, 4.1, 4.2, and 4.3 respectively.} \label{figure2}
\end{figure}
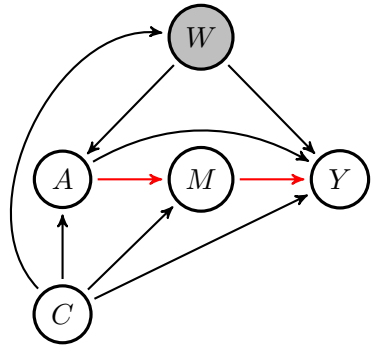
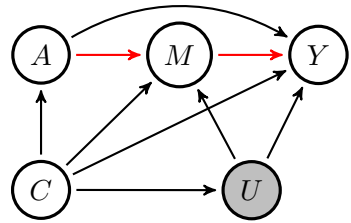
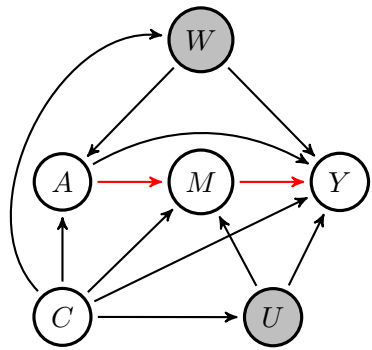
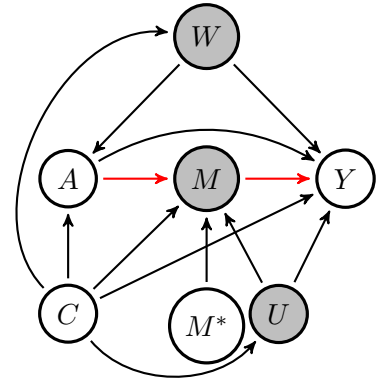

\section*{Proofs for all results} 

\subsection*{Result 1}

\noindent Under DAG (a), the true NIE (i.e. $W$ is observed) is given by,
\begin{align}
NIE_{truth}(a,a^*)  & \stackrel{M1-M4'}{=} \int_c \int_m \int_w E[Y | m, a, c, w] (f_M(m | a,c) - f_M(m \mid a^*,c)) f_W(w|c) f_C(c) dw dm dc \notag \\
& =\int_c  \int_m \int_w E[Y | m, a, c, w] (f_M(m | a,c) - f_M(m \mid a^*,c)) f_W(w|c) f_C(c) dw dm dc \notag \\
& = \int_c \int_m (f_M(m | a,c) - f_M(m \mid a^*,c)) \bigg\{ \int_w E[Y | m, a,c,w] f_W(w|c) dw \bigg\} f_C(c) dm dc \notag \\
& = \int_c \int_m (f_M(m | a,c) - f_M(m \mid a^*,c)) \bigg\{ \int_w (E[Y | m, a, c, w] \ - \notag \\
& \hspace{7cm} E[Y | M=0, a, c, w]) f_W(w|c) dw \bigg\}  f_C(c) dm dc \notag \\
& \stackrel{M4'(c)}{=} \int_c \int_m (f_M(m | a,c) - f_M(m \mid a^*,c)) \bigg\{ \int_w \gamma_1(a,m,c) f_W(w|c) dw \bigg\} f_C(c) dm dc \notag  \\
& = \int_c \int_m \gamma_1(a,m,c) (f_M(m | a, c) - f_M(m \mid a^*, c)) f_C(c) dm dc\label{Anie_identified_truth} \tag{A1}
\end{align}

\noindent The naive NIE which does not use information on $W$ and incorrectly assumes M1-M4 holds is given by, 
\begin{align}
NIE_{naive}(a,a^*)  & = \int_c \int_m E[Y | M=m, A=a, C=c] (f_M(m | a, c) - f_M(m \mid a^*, c)) dm dc\notag \\
& = \int_c \int_m \underbrace{(E[Y | M=m, A=a, C=c] - E[Y | M=0, A=a, C=c])}_\text{= $\gamma_1(a,m,c)$ by **} (f_M(m | a, c) - f_M(m \mid a^*, c)) dm \notag \\
& = NIE_{truth}(a,a^*)  \notag
\end{align}

\noindent **We want to show that $\gamma_1(a,m,c)$ can be identified without conditioning on $W$. To do so, we first recognize that the no additive interaction assumption encodes the following, 

\begin{align*}
E[Y | M=m, A=a, C=c, W=w] & = \theta_{A}(a) + \theta_{M}(m) + \theta_{C}(c) + \theta_{W}(w) \\
& \hspace{1.5cm} + \theta_{A,M}(a,m) + \theta_{A,C}(a,c) + \theta_{M,C}(m,c) + \theta_{A,W}(a,w) + \theta_{W,C}(w,c) \\
& \hspace{3cm} + \theta_{A,M,C}(a,m,c) + \theta_{A,W,C}(a,w,c)
\end{align*} 
\noindent so that 
$$\gamma_1(a,m,c) = \theta_M(m) + \theta_{A,M}(a,m) + \theta_{M,C}(m,c)  + \theta_{A,M,C}(a,m,c)$$
\noindent where $\theta_{A}(a)$, $\theta_{M}(m)$, and $\theta_{W}(w)$ are arbitrary functions of $a,m,$ and $w$, respectively, with the property that $\theta_{A}(0)=\theta_{M}(0)=\theta_{W}(0)=0$. Likewise, $\theta_{A,M}(a,m)$ and $\theta_{A,W}(a,w)$ are arbitrary functions of interactions between $A-M$ and $A-W$ such that $\theta_{A,M}(0,0)=\theta_{A,M}(0,m)=\theta_{A,M}(a,0)=0$ and $\theta_{A,W}(0,0)=\theta_{A,W}(0,w)=\theta_{A,W}(w,0)=0$. This idea can be extended to the other terms encoding interactions. 

The marginal model is given by,
\begin{align*}
E[Y | M=m, A=a, C=c]  & = E[  E[Y | M=m, A=a, C=c, W=w] \mid A=a,M=m,C=c] \\
& = \int_w E[Y | M=m, A=a, C=c, W=w]  f_W(w | a,m,c) dw \\
& = \int_w E[Y | M=m, A=a, C=c, W=w]  f_W(w | a,c) dw \\
& \stackrel{M4'}{=} \theta_{A}(a) + \theta_{M}(m) + \theta_{C}(c) + E[\theta_W(w) | A=a,C=c] \\
& \hspace{1cm} + \theta_{A,M}(a,m) + \theta_{A,C}(a,c) + \theta_{M,C}(m,c) + E[\theta_{A,W}(a,w)| A=a,C=c]   \\
& \hspace{1cm} + E[\theta_{W,C}(w,c)| A=a,C=c] + \theta_{A,M,C}(a,m,c) + E[\theta_{A,W,C}(a,w,c)| A=a,C=c]  
\end{align*}

\noindent So that the marginal contrast (i.e. $W$ not observed) is equal to,
\begin{align*}
E[Y | M=m, A=a,C=c] - E[Y | M=0, A=a,C=c] & = \theta_M(m) + \theta_{A,M}(a,m) + \theta_{M,C}(m,c)  + \theta_{A,M,C}(a,m,c) \\
& = \gamma_1(a,m,c)
\end{align*}

\subsection*{Result 2} 

\noindent Under DAG (a), the true $RR^{NIE}$ is (i.e. $W$ is observed),
\begin{align}
RR_{truth}^{NIE}(a,a^* | c,w) & \stackrel{M1-M4''}{=}  \frac{ \int_m Pr[Y=1 |a,M=m,w,c] f_M(m |a,c) \ dm }{ \int_m Pr[Y=1 |a,M=m,w,c] f_M(m |a^*,c)  \ dm }  \notag \\
& = \frac{ \int_m Pr[Y=1 | a,M=m,w,c] f_M(m |a,c ) \ dm }{ \int_m Pr[Y=1 |a,M=m,w,c] f_M(m |a^*,c)  \ dm }  \times \frac{P[Y=1 | a, M=0 , w,c]}{P[Y=1 | a, M=0 , w,c]} \notag \\
& = \frac{ \int_m Pr[Y=1 | a,M=m ,w,c] \big/ Pr[Y=1 | a,M=0 ,w,c]  f_M(m | a,c ) \ dm }{ \int_m Pr[Y=1 | a,M=m ,w,c] \big/ Pr[Y=1 | a,M=0 ,w,c]  f_M(m | a^*,c ) \ dm }  \notag \\
& \stackrel{M4''(c)}{=}   \frac{ \int_m  \gamma_2(a,m,c ) f_M(m | a,c ) \ dm }{ \int_m \gamma_2(a,m,c )  f_M(m | a^*,c ) \ dm }\label{Anie_identified3} \tag{A2}
\end{align}

\noindent The naive NIE which does not use information on $W$ and incorrectly assumes M1-M4 holds is given by, 
\begin{align}
RR_{naive}^{NIE}(a,a^* | c) & = \frac{ \int_m Pr[Y=1 |a,M=m,c] f_M(m |a,c) \ dm }{ \int_m Pr[Y=1 |a,M=m,c] f_M(m |a^*,c)  \ dm }  \notag \\
& = \frac{ \int_m Pr[Y=1 | a,M=m,c] f_M(m |a,c) \ dm }{ \int_m Pr[Y=1 |a,M=m,c] f_M(m |a^*,c)  \ dm }  \times \frac{P[Y=1 | a, M=0,c]}{P[Y=1 | a, M=0,c]} \notag \\
& = \frac{ \int_m Pr[Y=1 | a,M=m,c] \big/ Pr[Y=1 | a,M=0,c]  f_M(m | a,c) \ dm }{ \int_m Pr[Y=1 | a,M=m,c] \big/ Pr[Y=1 | a,M=0,c]  f_M(m | a^*,c) \ dm }  \notag \\
& =  \frac{ \int_m  \gamma_2(a,m,c) f_M(m | a,c) \ dm }{ \int_m \gamma_2(a,m,c)  f_M(m | a^*,c) \ dm } \tag{A3}
\end{align}

\noindent **We want to show that $\gamma_2(a,m,c)$ can be identified without conditioning on $W$. First note that M4$''$ encodes a no M-W interaction assumption; 

\begin{align*}
Pr(Y=1 | M=m, A=a, C=c, W=w) & = exp(\theta_{A}(a) + \theta_{M}(m) + \theta_{C}(c) + \theta_{W}(w) \\
& \hspace{1.5cm} + \theta_{A,M}(a,m) + \theta_{A,C}(a,c) + \theta_{M,C}(m,c) + \theta_{A,W}(a,w) + \theta_{W,C}(w,c) \\
& \hspace{3cm} + \theta_{A,M,C}(a,m,c) + \theta_{A,W,C}(a,w,c))
\end{align*} 

so that,

\begin{align*}
\gamma_2(a,m)  & = Pr[Y=1| M=m, A=a, C=c, W=w] / Pr[Y=1 | M=0, A=a, C=c, W=w] \\
& =  exp(\theta_M(m) + \theta_{A,M}(a,m) + \theta_{M,C}(m,c)  + \theta_{A,M,C}(a,m,c))
\end{align*}

The marginal model is given by,
\begin{align*}
E[Y | M=m, A=a,C=c]  & = E[  E[Y | M=m, A=a, C=c, W=w] \mid A=a,M=m,C=c] \\
& = \int_w E[Y | M=m, A=a, W=w, C=c]  f_W(w | a,m,c) dw \\
& = \int_w E[Y | M=m, A=a, W=w, C=c]  f_W(w | a,c) dw \\
& = \int_w Pr[Y =1 | M=m, A=a, W=w, C=c]  f_W(w | a,c) dw \\
& = \int_w exp(\theta_{A}(a) + \theta_{M}(m) + \theta_{C}(c) + \theta_{W}(w) + \theta_{A,M}(a,m) + \theta_{A,C}(a,c) + \theta_{M,C}(m,c) \\
& \hspace{1.5cm} + \theta_{A,W}(a,w) + \theta_{W,C}(w,c) + \theta_{A,M,C}(a,m,c) + \theta_{A,W,C}(a,w,c)) f_W(w | a,c) dw  \\
& = \exp( \theta_A(a) + \theta_M(m) + \theta_{C}(c) + \theta_{A,M}(a,m) + \theta_{A,C}(a,c) + \theta_{M,C}(m,c) + \theta_{A,M,C}(a,m,c) ) \\
& \hspace{1.5cm} \times E[ exp(\theta_W(w) + \theta_{A,W}(a,w) + \theta_{W,C}(w,c) + \theta_{A,W,C}(a,w,c)) | A=a, C=c ]
\end{align*}

\noindent So that the marginal ratio (i.e. $W$ not observed) is equal to,
\begin{align*}
E[Y | M=m, A=a,C=c]/E[Y | M=0, A=a,C=c] & = \exp(\theta_M(m) + \theta_{A,M}(a,m) + \theta_{M,C}(m,c) + \theta_{A,M,C}(a,m,c) ) \\
& = \gamma_2(a,m,c)
\end{align*}

\subsection*{Hazard difference natural indirect effect proof} 

Denote the hazard difference natural indirect effect as the following, 

$$ HD^{NIE} = \lambda_{T_{a,M_{a}}}(t) - \lambda_{T_{a,M_{a^*}}}(t) $$

\noindent Define the following condition, 
$$\textrm{\textbf{M4$'''$}} \ \ \frac{Pr(T > t |a,m,w,c)}{Pr(T > t |a,m,w,c)}= \exp \big\{- \int_0^t \lambda_T(u | a,m,w,c) - \lambda_T(u | a,m=0,w,c) du \big\} = \gamma_3(a,m,c,t)$$
\noindent equivalently, 
$$\textrm{\textbf{M4$'''$}} \ \ \lambda_T(t | a,m,w,c) - \lambda_T(t | a,m=0,w,c) = \gamma_3'(a,m,c,t)$$
\noindent where 
\begin{align*}
\lambda_T(t | a,m,w,c) & = \lambda_0(t) + \lambda_A(t)a + \lambda_M(t)m + \lambda_{A,M}(t)am \\
& + \lambda_{C}(t)c + \lambda_{A,C}(t)ac + \lambda_{A,C}(t)mc + \lambda_{A,M,C}(t)amc \\
& + \lambda_{W,C}(t)wc + \lambda_{W,A}(t)wa + \lambda_{W,A}(t)wac 
\end{align*}  

\noindent  Note that M4$'''$ encodes a no multiplicative interaction on the survival probability scale or a no additive interaction on the hazard difference scale, that is, 
\begin{align*}
\lambda_T(t | a,m,c,w) - \lambda_T(t | a,m=0,c,w) & = \gamma_3'(a,m,c,t)  \\
& =  \lambda_{M}(t)m + \lambda_{A,M}(t)am + \lambda_{A,C}(t)ac + \lambda_{M,C}(t)mc+ \lambda_{A,M,C}(t)amc  \\
\end{align*}

\noindent We show that under assumptions M1-M3, M4$'''$, and M5, the conditional $HD^{NIE}$ is identified. Under Figure 2a, the true conditional $HD^{NIE}$ (i.e. $W$ is observed) on the survival scale is as follows. 
\begin{align*}
\frac{Pr(T_{a,M_a} > t)}{Pr(T_{a,M_{a^*}} > t)} & = \frac{ \int_m Pr(T > t | A=a, M=m, C=c, W=w) Pr(M=m | A=a,C=c) dm }{\int_m Pr(T > t | A=a, M=m, C=c, W=w) Pr(M=m | A=a^*,C=c) dm } \\
& = \frac{ \int_m exp( - \int_0^t  \lambda_T(t | a,m,c,w)  du ) Pr(M=m | A=a,C=c) dm }{\int_m  exp( - \int_0^t  \lambda_T(t | a,m,c,w)  du ) Pr(M=m | A=a^*,C=c) dm} \\ 
& = \frac{ \int_m exp( - \int_0^t \lambda_T(u | a,m,c,w)   du ) Pr(M=m | A=a,C=c) dm }{\int_m  exp( - \int_0^t  \lambda_T(u | a,m,c,w)  du ) Pr(M=m | A=a^*,C=c) dm} \frac{exp(\int_0^t \lambda_T(u | a,m=0,c,w)   du )}{exp(\int_0^t \lambda_T(u | a,m=0,c,w)   du )} \\ 
& = \frac{ \int_m \gamma_3(a,m,c,t)  Pr(M=m | A=a,C=c) dm }{\int_m \gamma_3(a,m,c,t) Pr(M=m | A=a^*,C=c) dm} \\ 
\end{align*}
\noindent The naive $HD^{NIE}$ on the survival scale which does not use information on $W$ is given by, 
\begin{align*}
\frac{Pr(T_{a,M_a} > t)}{Pr(T_{a,M_{a^*}} > t)} & = \frac{ \int_m Pr(T > t | A=a, M=m, C=c) Pr(M=m | A=a,C=c) dm }{\int_m Pr(T > t | A=a, M=m, C=c) Pr(M=m | A=a^*,C=c) dm } \\
& = \frac{ \int_m exp( - \int_0^t  \lambda_T(t | a,m,c)  du ) Pr(M=m | A=a,C=c) dm }{\int_m  exp( - \int_0^t  \lambda_T(t | a,m,c)  du ) Pr(M=m | A=a^*,C=c) dm} \\ 
& = \frac{ \int_m exp( - \int_0^t \lambda_T(u | a,m,c)   du ) Pr(M=m | A=a,C=c) dm }{\int_m  exp( - \int_0^t  \lambda_T(u | a,m,c)  du ) Pr(M=m | A=a^*,C=c) dm} \frac{exp(\int_0^t \lambda_T(u | a,m=0,c)   du )}{exp(\int_0^t \lambda_T(u | a,m=0,c)   du )} \\ 
& = \frac{ \int_m \gamma_3(a,m,c,t)  Pr(M=m | A=a,C=c) dm }{\int_m \gamma_3(a,m,c,t) Pr(M=m | A=a^*,C=c) dm} 
\end{align*}
\noindent **We want to show that $\gamma_3(a,m,c)$ can be identified without conditioning on $W$. The marginal model is given by,
\begin{align*} 
Pr(T > t | A=a, M=m, C=c, W=w) &  = \int_w Pr(T > t | a, m, c, w) f(w | a, m, c)  dw\\
& = \int_w Pr(T > t | a, m, c, w) f(w | a, c) dw\\
& = \int_w \exp \bigg\{ - \int_0^t  \lambda_T(t | a,m,c,w)  du \bigg\} f(w | a, c) dw\\
& = \int_w \exp \bigg\{  - \int_0^t  \lambda_0(u) + \lambda_A(u)a + \lambda_M(u)m + \lambda_{A,M}(u)am \\
& \hspace{1cm} + \lambda_{C}(u)c + \lambda_{A,C}(u)ac + \lambda_{A,C}(u)mc + \lambda_{A,M,C}(u)amc \\
& \hspace{1cm} + \lambda_{W,C}(u)wc + \lambda_{W,A}(u)wa + \lambda_{W,A}(u)wac  du \bigg\}  f(w | a, c) dw\\
& = \exp \bigg\{ - \int_0^t  \lambda_0(u) + \lambda_A(u)a + \lambda_M(u)m + \lambda_{A,M}(u)am \\
& \hspace{1cm} + \lambda_{C}(u)c + \lambda_{A,C}(u)ac + \lambda_{A,C}(u)mc + \lambda_{A,M,C}(u)amc du \bigg\} \\
& \hspace{1cm} \times E[\exp \big\{- \int_0^t \lambda_{W,C}(u)wc + \lambda_{W,A}(u)wa \\
& \hspace{4cm} + \lambda_{W,A}(u)wac \ du \big\} | A=a, C=c]   
\end{align*}
\noindent So that the marginal survival probability ratio (i.e. $W$ not observed) is equal to,
$$ \frac{Pr(T > t | A=a, M=m, C=c, W=w)}{Pr(T > t | A=a, M=m, C=c, W=w)} = \gamma_3(a,m,c,t) $$
\noindent and the hazard difference natural indirect effect (i.e. $W$ not observed) is equal to, 
\begin{align*} 
\lambda_T(t | a,m,c) - \lambda_T(t | a,m=0,c) & = \frac{d}{dt} \big\{ \log (Pr(T > t | A=a, M=m, C=c) ) \\
& \hspace{2cm} - \log (Pr(T > t | A=a, M=0, C=c) ) \big\} \\
& = \gamma_3'(a,m,c,t)
\end{align*}

\subsection*{Hazard ratio natural indirect effect proof} 

\noindent Define the following conditions, 
$$\textrm{\textbf{M4$''''$}} \ \ \lambda_T(t | a,m,w,c)/\lambda_T(t | a,m=0,w,c) = \gamma_4(a,m,c) $$
$$\textrm{\textbf{M5}} \ \ exp(- \Lambda_t) = 1 \ \textrm{ (rare outcome) } $$

\noindent  Note that M4$''''$ encodes a no multiplicative interaction on the hazard ratio scale, that is, 

\begin{align*}
\lambda_T(t | a,m,c,w) & =  \lambda_0(t) \exp(\theta_{A}(a) + \theta_{M}(m) + \theta_{C}(c) + \theta_{W}(w) \\
& \hspace{1.5cm} + \theta_{A,M}(a,m) + \theta_{A,C}(a,c) + \theta_{M,C}(m,c) + \theta_{A,W}(a,w) + \theta_{W,C}(w,c) \\
& \hspace{3cm} + \theta_{A,M,C}(a,m,c) + \theta_{A,W,C}(a,w,c))  \\
\lambda_T(t | a,m,c,w)/\lambda_T(t | a,m=0,c,w) & = \gamma_4(a,m,c)  \\
& =  \exp(\theta_{M}(m) + \theta_{A,M}(a,m) + \theta_{M,C}(m,c) + \theta_{A,M,C}(a,m,c)) \\
\end{align*}

\noindent We show that under assumptions M1-M3, M4$''''$, and M5, the conditional $HR^{NIE}$ is identified. 

\noindent Under Figure 2a, the true $HR^{NIE}$ is (i.e. $W$ is observed),
\begin{align*}
HR_{truth}^{NIE}(a,a^*|c) & = \lambda_{T_{aM_a}}(t | a,m,c,w) /  \lambda_{T_{aM_{a^*}}}(t | a,m,c,w) \\
& = \frac{ f_{T_{aM_a}}(t | a,m,c,w) S_{T_{aM_{a^*}}}(t | a,m,c,w) }{ f_{T_{aM_{a^*}}}(t | a,m,c,w) S_{T_{aM_{a}}}(t |a,m,c,w) } \\
& = \frac{\int_m f_{T}(t | a, m,c, w) f_{M}(m | a)  \int_m S_{T}(t | a, m,c, w) f_{M}(m | a^*) }{\int_m f_{T}(t | a, m, c,w) f_{M}(m | a^*,c)  \int_m S_{T}(t | a, m, c,w) f_{M}(m | a,c) } \\
& = \frac{\int_m \lambda_{T}(t | a, m,c, w)  S_{T}(t | a, m,c, w) f_{M}(m | a,c)  \int_m S_{T}(t | a, m,c, w) f_{M}(m | a^*,c) }{\int_m \lambda_{T}(t | a, m,c, w)  S_{T}(t | a, m,c, w)  f_{M}(m | a^*,c)  \int_m S_{T}(t | a, m,c, w) f_{M}(m | a,c) } \\
& = \frac{\int_m \lambda_{T}(t | a, m,c, w)  exp(-\Lambda_t) f_{M}(m | a,c)  \int_m exp(-\Lambda_t) f_{M}(m | a^*,c) }{\int_m \lambda_{T}(t | a, m,c, w)  exp(-\Lambda_t)  f_{M}(m | a^*,c)  \int_m exp(-\Lambda_t) f_{M}(m | a,c) } \\
& \stackrel{M6}{\approx} \frac{\int_m \lambda_{T}(t | a, m,c, w)  f_{M}(m | a,c) dm  \int_m  f_{M}(m | a^*,c) dm }{\int_m \lambda_{T}(t | a, m,c, w)  f_{M}(m | a^*,c) dm  \int_m  f_{M}(m | a,c) dm }  \\
& =\frac{\int_m \lambda_{T}(t | a, m,c, w)  f_{M}(m | a,c) dm }{\int_m \lambda_{T}(t | a, m,c, w)  f_{M}(m | a^*,c) dm}  \\
& =\frac{\int_m \lambda_{T}(t | a, m,c, w)  f_{M}(m | a,c) dm }{\int_m \lambda_{T}(t | a, m,c, w)  f_{M}(m | a^*,c) dm}  \frac{\lambda_{T}(t | a, m=0, w)}{\lambda_{T}(t | a, m=0,c, w)}\\
& = \frac{\int_m \gamma_4(a,m) f_{M}(m | a,c) dm }{\int_m \gamma_4(a,m,c)  f_{M}(m | a^*,c) dm} \\
\end{align*}

\noindent The naive $HR^{NIE}$ which does not use information on $W$ is given by, 
\begin{align*}
HR_{naive}^{NIE}(a,a^*|c) & = \lambda_{T_{aM_a}}(t) /  \lambda_{T_{aM_{a^*}}}(t) \\
& = \frac{ f_{T_{aM_a}}(t | a,m,c) S_{T_{aM_{a^*}}}(t| a,m,c) }{ f_{T_{aM_{a^*}}}(t| a,m,c) S_{T_{aM_{a}}}(t| a,m,c) } \\
& = \frac{\int_m f_{T}(t | a, m,c) f_{M}(m | a,c)  \int_m S_{T}(t | a, m,c) f_{M}(m | a^*,c) }{\int_m f_{T}(t | a, m,c) f_{M}(m | a^*,c)  \int_m S_{T}(t | a, m,c) f_{M}(m | a,c) } \\
& = \frac{\int_m \lambda_{T}(t | a, m,c)  S_{T}(t | a, m,c) f_{M}(m | a,c)  \int_m S_{T}(t | a, m,c) f_{M}(m | a^*,c) }{\int_m \lambda_{T}(t | a, m,c)  S_{T}(t | a, m,c)  f_{M}(m | a^*,c)  \int_m S_{T}(t | a, m,c) f_{M}(m | a,c) } \\
& = \frac{\int_m \lambda_{T}(t | a, m,c)  exp(-\Lambda_t) f_{M}(m | a,c)  \int_m exp(-\Lambda_t) f_{M}(m | a^*,c) }{\int_m \lambda_{T}(t | a, m,c)  exp(-\Lambda_t)  f_{M}(m | a^*,c)  \int_m exp(-\Lambda_t) f_{M}(m | a,c) } \\
& \stackrel{M6}{\approx} \frac{\int_m \lambda_{T}(t | a, m,c)  f_{M}(m | a,c) dm  \int_m  f_{M}(m | a^*,c) dm }{\int_m \lambda_{T}(t | a, m,c)  f_{M}(m | a^*,c) dm  \int_m  f_{M}(m | a,c) dm }  \\
& =\frac{\int_m \lambda_{T}(t | a, m,c)  f_{M}(m | a,c) dm }{\int_m \lambda_{T}(t | a, m,c)  f_{M}(m | a^*,c) dm}  \\
& =\frac{\int_m \lambda_{T}(t | a, m,c)  f_{M}(m | a,c) dm }{\int_m \lambda_{T}(t | a, m,c)  f_{M}(m | a^*,c) dm}  \frac{\lambda_{T}(t | a, m=0,c)}{\lambda_{T}(t | a, m=0,c)}\\
& = \frac{\int_m \gamma_4(a,m,c) f_{M}(m | a,c) dm }{\int_m \gamma_4(a,m,c)  f_{M}(m | a^*,c) dm} \\
\end{align*}
\noindent **We want to show that $\gamma_4(a,m,c)$ can be identified without conditioning on $W$. The marginal model is given by,
\begin{align*} 
\lambda_T(t|a,m,c) &  \approx f_t(t|a,m,c) \\
& = \int_w f_t(t|a,m,w,c) f(w|a,m,c) dw \\
& \approx \int_w \lambda_t(t|a,m,w,c) f(w|a,m,c) dw \\
& = \int_w  \lambda_t(t|a,m,w,c) f(w|a,c) dw \\
& = \int_w \lambda_0 \exp \big(\theta_{A}(a) + \theta_{M}(m) + \theta_{C}(c) + \theta_{W}(w) \\
& \hspace{1.5cm} + \theta_{A,M}(a,m) + \theta_{A,C}(a,c) + \theta_{M,C}(m,c) + \theta_{A,W}(a,w) + \theta_{W,C}(w,c) \\
& \hspace{3cm} + \theta_{A,M,C}(a,m,c) + \theta_{A,W,C}(a,w,c) \big)  f_W(w | a,c) dw \\
& = \lambda_0 \exp( \theta_A(a) + \theta_M(m) + \theta_{C}(c) + \theta_{A,M}(a,m) + \theta_{A,C}(a,c) + \theta_{M,C}(m,c) + \theta_{A,M,C}(a,m,c) ) \\
& \hspace{1.5cm} \times E[ exp(\theta_W(w) + \theta_{A,W}(a,w) + \theta_{W,C}(w,c) + \theta_{A,W,C}(a,w,c)) | A=a, C=c ]
\end{align*}
\noindent So that the marginal ratio (i.e. $W$ not observed) is equal to,
\begin{align*} 
\lambda_T(t|a,m,c)/\lambda_T(t|a,m=0,c) & = \exp(\theta_M(m) + \theta_{A,M}(a,m) + \theta_{M,C}(m,c) + \theta_{A,M,C}(a,m,c) )\\
& = \gamma_4(a,m,c)
\end{align*}

\subsection*{Result 3} 

M3$'$(c) and (d) imply the following modeling assumptions:
\begin{align}
E[Y | A, M, C, U] = \theta_0 + \theta_m M + \theta_a (A) + \theta_c^T(C)  + \theta_u (U) \label{Ay_genius} \tag{A4} \\
E[M | A, C, U] = \beta_0 + \beta_a A + \beta_c(C) + \beta_u(U) \label{Am_genius} \tag{A5}
\end{align} 
\noindent Note that interactions between $A,C$ and $M,C$ are allowed. This is omitted from the above equations for simplicity, but the proof would proceed similarly.

\noindent Under DAG (b), the true NIE (i.e. $U$ is observed) is given by,
\begin{align}
NIE_{truth}(a,a^*)  & \stackrel{M1-M3'}{=}  \int_{c,m,u} E[Y | m, a, c, u] (f_M(m | a, c, u) - f_M(m \mid a^*, c, u)) dm df_U(u|c) df_C(c) \notag \\
& = \int_{m,u} E[Y | m, a, c, u] (f_M(m | a, c, u) - f_M(m \mid a^*, c, u)) dm df_U(u|c) df_C(c) \notag \\
& \stackrel{eq \ref{Ay_genius}}{=} \int_{c,m,u} (f_M(m | a, c, u) - f_M(m \mid a^*, c, u)) (\theta_m M + \theta_a (A) + \theta_c^T (C) + \theta_u (U)) dmdf_U(u|c) df_C(c) \notag \\
& = \int_{c,u}  E[M | a,c,u] - E[M | a^*,c, u] df_U(u|c) df_C(c) \notag \\
& \stackrel{eq \ref{Am_genius}}{=} \theta_m \beta_a (a - a^*)  \notag
\end{align}

\noindent Now, it suffices to show that we can find unbiased estimators for $\theta_m$ and $\beta_a$. The former follows directly from GENIUS. Under the conditions given in assumption M3$'$, 

\begin{align}
\theta_m & = \frac{E\left[\{A-E[A|C]\}\{M-E[M|  A,C]\}Y\right]}{E\left[\{A-E[A,C]\}\{M-E[M| A,C]\}M\right]} \notag \\
& =\frac{E\left[\{A-E[A|C]\}\{M-E[M|  A,C]\}Y\right]}{
	\text{var(A)}[\text{var}(M|A=1,C)-\text{var}(M|A=0,C)]
} \notag
\end{align}

\noindent The latter follows from the fact that $A \perp U | C=c \ \forall c$,

\begin{align*}
E[M | A=a, C=c] & = E[E[M | A=a,C=c,U=u] | A=a, C=c] \\
& \stackrel{eq \ref{Am_genius}}{=} \int_u (\beta_0 + \beta_a a + \beta_c(c) + \beta_u(U))  f(u | a,c)   \\
& \stackrel{M3'(b)}{=} \int_u (\beta_0 + \beta_a a + \beta_c(c) + \beta_u(U))   f(u|c) \\
& = \beta_0 + \beta_a a + \beta_c(c) + E[\beta_u(U) | C=c] \\
& = \beta_0^* + \beta_a a + \beta_c(c)^* c
\end{align*}

\noindent That is, $\beta_a$ can be consistently estimated by the OLS estimator, $\hat{\beta}_a$ from the model $E[M | A=a, C=c] = \beta_0^* + \beta_a a + \beta_c(c)^* c$. 

\subsection*{Result 4} 

M3$''$(f) implies the following new modeling assumption for the outcome model:
\begin{align}
E[Y | A, M, C, U] = \theta_0 + \theta_m M + \theta_a (A) + \theta_c^T(C)  + \theta_u (U) + \theta_w (W) \label{Ay_genius2} \tag{A6} 
\end{align} 

This proof follows similarly to that of Result 3. Under DAG (c), the true NIE (i.e. $U$, $W$ are observed) is given by,
\begin{align}
NIE_{truth}(a,a^*)  & \stackrel{M1-M3''}{=}  \int_{m,u,w,c} E[Y |m, a, c, u, w] (f_M(m | a, c,u) - f_M(m \mid a^*,c, u)) dm df_W(w|c) df_U(u|c) df_C(c)  \notag \\
& = \int_{c,m,u} (f_M(m | a,c,u) - f_M(m \mid a^*,c,u))  \notag \\
& \hspace{1cm} \times \bigg\{ \int_w E[Y | M=m, A=a,C=c,U=u,W=w] f_W(w) dw \bigg\}dm df_W(w|c) df_U(u|c) df_C(c)  \notag \\
& = \int_{c,m,u} (f_M(m | a,c,u ) - f_M(m \mid a^*,c,u)) \bigg\{ \int_w (E[Y | M=m, A=a,C=c U=u,W=w] \ - \notag \\
& \hspace{2cm} E[Y | M=0, A=a, C=c,U=u,W=w]) f_W(w|c) dw \bigg\}  dmdf_U(u|c) df_C(c)  \notag \\
& \stackrel{eq \ref{y_genius2}}{=} \theta_m \int_{m,u} m (f_M(m | a,c,u ) - f_M(m \mid a^*,c,u)) dm df_U(u|c) df_C(c) \notag \\
& \stackrel{eq \ref{Am_genius}}{=} \theta_m \beta_a (a-a^*) \notag
\end{align}
\noindent Similar to the proof of Result 3, it suffices to show that we can find unbiased estimators for $\theta_m$ and $\beta_a$. For the latter, a slight update is made to the GENIUS proof, 
\begin{align}
E[ (A - E(A|C)) (M - E(M | A,C)) Y] &  = E[  E[ (A - E[A|C]) (M - E[M | A|C]) Y | A, M, C, U, W] ] \notag \\
& = E[ (A - E[A|C])) (M - E[M | A|C]) E[Y | A, M, C, U, W] ]  \notag \\
& \stackrel{eq \ref{y_genius2}}{=} E[ (A - E[A|C])) (M - E[M | A, C]) \{ \theta_0 + \theta_m M + \theta_a (A) \notag \\
& \hspace{3cm} + \theta_c (C) + \theta_u (U) + \theta_w (W) \} ]  \notag \\
& = E[ (A - E[A|C])) (M - E[M | A,C]) \theta_m M  ] \label{genius1} \tag{A7} \\
& \hspace{.5cm} + E[ (A - E[A|C])) (M - E[M | A,C]) \theta_a (A) ] \label{genius2} \tag{A8} \\
& \hspace{.5cm} + E[ (A - E[A|C])) (M - E[M | A,C]) \theta_c (C) ] \label{genius3} \tag{A9} \\
& \hspace{.5cm} + E[ (A - E[A|C])) (M - E[M | A,C]) \theta_u (U) ] \label{genius4} \tag{A10} \\
& \hspace{.5cm} + E[ (A - E[A|C])) (M - E[M | A,C]) \theta_w (W) ] \label{genius5} \tag{A11} \\
& = E[ (A - E[A|C])) (M - E[M | A,C]) \theta_m M  ] \notag \\
& \implies \theta_m =\frac{ E[ (A - E(A|C)) (M - E(M | A,C)) Y] }{ E[ (A - E[A|C])) (M - E[M | A,C]) M] } \notag
\end{align}
\noindent The terms (\ref{genius1}) - (\ref{genius4}) are identical to the GENIUS proof. That is, (\ref{genius2})-(\ref{genius4}) are identically zero. Here we show (\ref{genius5}) is identically zero: 
\begin{align*}
(4) & = E[ (A - E[A|C])) (M - E[M | A,C]) \theta_w (W) ]  \\
& = E[ E[ (A - E[A|C])) (M - E[M | A,C]) \theta_w (W) | A,C]  ] \\
& = E[ (A-E[A|C])   E[ (M - E[M | A,C]) \theta_w (W) | A,C ] ] \\
& = E[ (A-E[A|C])   \underbrace{E[ M - E[M | A,C] | A,C  ]}_{=0} E[\theta_w (W) | A,C] \\
& = 0 
\end{align*}
\noindent Thus, we recover the GENIUS identifying formula for $\theta_m$. For $\beta_a$, the fact that $A \perp U | C$,
\begin{align*}
E[M | A=a,C=c] & = E[E[M | A=a,C=c,U=u] | A=a] \\
& \stackrel{eq \ref{Am_genius}}{=} \int_u (\beta_0 + \beta_a a + \beta_c(c) + \beta_u(U)) f(u | a,c) \\
& \stackrel{M3'''(b)}{=} \int_u (\beta_0 + \beta_a a + \beta_c(c) + \beta_u(U)) f(u|c) \\
& = \beta_0 + \beta_a a + \beta_c(c) + E[\beta_u(U)|C=c] \\
& = \beta_0^* + \beta_a a + \beta_c^*(c)
\end{align*}
\noindent That is, $\beta_a$ can be consistently estimated by the OLS estimator, $\hat{\beta}_a$ from the model $E[M | A=a] = \beta_0^* + \beta_a a +\beta_c^*(c)  $. 

\subsection*{Result 5} 
The proof follows from Result 4. Under DAG (d), the true NIE is given by, 
$$ NIE_{truth}(a,a^*)  = \theta_m \beta_a (a-a^*) $$ 
where $\theta_m$ comes from the model in equation (\ref{Ay_genius2}) and $\beta_a$ comes from the model in (\ref{Am_genius}). It suffices to show that we can find unbiased estimators for $\theta_m$ and $\beta_a$ even in the presence of measurement error.

Recall the following from the proof in Result 4 it follows that, 
\begin{equation}\label{combine}
E[ (A - E(A|C)) (M - E(M | A|C)) Y] =E[ (A - E(A|C)) (M - E(M | A,C)) \beta_mM] \tag{A12}
\end{equation}

\begin{equation*}\begin{split}
&E\left[\{A-E[A|C]\}\{M^*-E[M^*|  A,C]\}Y\right]\\
\stackrel{(E1-3)}{=}&E\left[\{A-E[A,C]\}\{M-E[M|  A,C]\}Y\right]+E\left[\{A-E[A|C]\}\{\epsilon-E[\epsilon|  A,C]\}Y\right]\\
=&E\left[\{A-E[A|C]\}\{M-E[M|  A,C]\}Y\right]+E\left[\{A-E[A|C]\}Y\right]\cdot E\{\epsilon-E[\epsilon]\}\\
=&E\left[\{A-E[A|C]\}\{M-E[M|  A,C]\}Y\right]\\
\stackrel{(\ref{combine})}{=}&E\left[\{A-E[A|C]\}\{M-E[M|  A,C]\}\theta_mM\right]\\
=&E\left[\{A-E[A|C]\}\{M-E[M|  A,C]\}\theta_mM\right]+E\left[\{A-E[A|C]\}\theta_mM\right]\cdot E\{\epsilon-E[\epsilon]\}\\
\stackrel{(E1-3)}{=}&E\left[\{A-E[A|C]\}\{M^*-E[M^*|  A,C]\}\theta_mM\right]\\
=&E\left[\{A-E[A|C]\}\{M^*-E[M^*|  A,C]\}\theta_mM^*\right].
\end{split}\end{equation*}
Therefore
\begin{equation*}\begin{split}
\theta_m = &\frac{E\left[\{A-E[A|C]\}\{M^*-E[M^*|  A,C]\}Y\right]}{E\left[\{A-E[A|C]\}\{M^*-E[M^*|  A,C]\}M^*\right]}\\
=&\frac{E\left[\{A-E[A|C]\}\{M^*-E[M^*|  A,C]\}Y\right]}{
	\text{var(A)}[\text{var}(M^*|A=1,C)-\text{var}(M^*|A=0,C)]
}
\end{split}\end{equation*}
\noindent We recover the MR GENIUS identifying formula for $\theta_m$. For $\beta_a$, the fact that $A \perp U | C$,
\begin{align*}
E[M^* | A=a,C=c] & = E[E[M^* | A=a,C=c,U=u] | A=a,C=c] \\
& \stackrel{E1}{=} E[E[M | A=a,C=c,U=u] | A=a,C=c] + E[\epsilon | A=a, C=c, U=u]\\
& \stackrel{E2}{=} E[E[M | A=a,C=c,U=u] | A=a,C=c] + E[\epsilon]\\
& \stackrel{E3}{=} E[E[M | A=a,C=c,U=u] | A=a,C=c] \\
& \stackrel{eq\ref{Am_genius}}{=} \int_u (\beta_0 + \beta_a a + \beta_c(c) + \beta_u(U)) f(u | a, c) \\
& \stackrel{M3'''(b)}{=} \int_u (\beta_0 + \beta_a a + \beta_c(c) + \beta_u(U)) f(u | c) \\
& = \beta_0 + \beta_a a + \beta_c(c) + E[\beta_u(U) | C=c] \\
& = \beta_0^* + \beta_a a + \beta_c^*(c)
\end{align*}
\noindent That is, $\beta_a$ can be consistently estimated by the OLS estimator, $\hat{\beta}_a$ from the model $E[M^* | A=a, C=c] = \beta_0^* + \beta_a a + \beta_c^*(c)$.

\section*{General Result 3 (incorporate interaction terms)}

We have noted in the above proofs and main text that assumptions M3$'$, M3$''$, and M3$'''$ can be made more general. We only show M3$'$ and Result 3 here, but the same result applies, \\
\textbf{M3$'$.}  There exists an unmeasured variable $U$ such that, 

\hspace{1cm}  (a) $ Y(a,m) \perp M(a^*) \mid C=c, U=u \ \forall \ a,a^*,c,u $ 

\hspace{1cm}  (b) $A \perp U \mid C=c$ $\forall \ c$ 

\hspace{1cm}  (c) There is no additive $M$--($U,A$) interaction in the model for $E[Y | A, M, C, U]$ 

$$E[Y | A, M=m, C, U] - E[Y | A, M=0, C, U] = \theta_m m + \theta_{mc} mc $$

\hspace{1.7cm} and no additive $A$--($U,C$) interaction in the model for $E[Y | A, M, C, U]$ 

$$E[Y | A=a, M, C, U] - E[Y | A=0, M, C, U] = \theta_a(a) + \theta_{ac}(a,c)$$

\hspace{1.7cm} for unknown functions $\theta_a(.)$ that satisfies $\theta_a(0) =0$ and $\theta_{ac}(.)$ that satisfies $\theta_{ac}(0,c) =0$.

\hspace{1cm}  (d) There is no additive $A$--($U$) interaction in the model for $E[M | A, C, U]$ 

$$E[M | A=a, C, U] - E[Y | A=0, C, U] = \beta_a (a) + \beta_{ac} (ac)$$

\hspace{1.7cm} for unknown functions $\beta_a(.)$ that satisfies $\beta_a(0) =0$ and $\beta_{ac}(.)$ that satisfies $\beta_{ac}(0,c) =0$.

\hspace{1cm}  (e) $\text{var}(M|A=a,C=c)-\text{var}(M|A=a',C=c)\neq 0 \ \ \forall \ a, a', c$

\noindent Under this more general assumption, Result 3 will become, 
\setcounter{theorem}{2}
\begin{theorem}
	Under assumptions M1, M2, and M3$'$, the natural indirect effect is uniquely identified by 
	\begin{align*}
	NIE(a,a^*) & = \theta_m \big\{ (\beta_a(a)  - \beta_a(a^*)) + E[\beta_{ac}(aC) - \beta_{ac}(a^*C)] \big\} \\
	& \hspace{1cm} + \theta_{mc}\big\{ (\beta_a(a)  - \beta_a(a^*)) E[C] + E[(\beta_{ac}(ac) - \beta_{ac}(a^*c))C] \big\} 
	\end{align*}  
\end{theorem}
\noindent Estimation of $\theta_m$ and $\theta_mc$ now proceeds from Section 4.4 where equation (9) becomes: 
\begin{align*}
0 & = n^{-1} \sum_{i=1}^n \bigg[ h(C_i) \big\{ A_i - \hat{E}[A | C_i] \big\} \big\{ M_i - \hat{E}[M | A_i, C_i] \big\} \big\{ Y_i - \hat{\theta}_mM_i - \hat{\theta}_{mc}M_iC_i \big\} \bigg] \label{mrgenius1_hat}
\end{align*}

\newpage
\bibliography{mybib}
\end{document}